# Molecular Doping of Electrochemically Prepared Triazine-based Carbon Nitride by 2,4,6-Triaminopyrimidine for Improved Photocatalytic Properties

*Leonard Heymann[1], Sophia Bittinger[1], Christian Klinke[1,2,\*]*

[1] Institute of Physical Chemistry, University of Hamburg,
Martin-Luther-King-Platz 6, 20146 Hamburg

[2] Chemistry Department, Swansea University - Singleton Park,
Swansea SA2 8PP, United Kingdom

*Corresponding author: christian.klinke@swansea.ac.uk

**Abstract** The copolymerization of melamine with 2,4,6-triaminopyrimidine (TAP) in an electrochemical induced polymerization process leads to the formation of molecular doped poly (triazine imide) (PTI). The polymerization is based on the electrolysis of water and evolving radicals during this process. The incorporation of TAP is shown by techniques such as elemental analysis, FTIR and NMR spectroscopy and powder x-ray diffraction (XRD), and it is shown that the carbon content can be tuned by the variation of the molar ratio of the two precursors. This incorporation of TAP directly influences the electronic structure of PTI and as a result a red-shift can be observed in UV-Vis spectroscopy. The smaller bandgap and the increased absorption in the visible range lead to improved photocatalytic properties. In dye degradation experiments it was possible to observe an increase of the rate of the degradation of methylene blue (MB) by a factor of 4 in comparison to undoped PTI or 7 if compared to melon.

## Introduction

Carbon nitrides as non-toxic and metal-free photocatalysts have been of particular interest in the scientific community starting latest with the discovery of photocatalytic water splitting by Wang et al. in 2009.[1] The different degrees of polymerization and their either triazine or heptazine based structure can be obtained by different synthesis procedures. These synthesis procedures lead to carbon nitride materials like melon (often referred as g-$C_3N_4$),[2–5] poly (triazine imide) (PTI),[6–9] poly (heptazine imide) (PHI),[10–12] and triazine-based graphitic carbon nitride (TGCN).[13–15]

Different paths were followed to improve the photocatalytic properties of these carbon nitrides. One example is to increase the specific surface and to modify the morphology to increase the number of accessible active sites. This can be done by exfoliation or by soft or hard template methods, which often results in improved photocatalytic properties.[16–18] Doping can be divided into molecular and elemental doping. Molecular doping describes the incorporation of different molecules, e.g. barbituric acid and 2,4,6-triaminopyrimidine (TAP).[19–23] Another promising molecule for molecular doping is 2-Aminobenzonitrile, which was used by Zhang et al. to dope g-$C_3N_4$. This doped carbon nitride has an increased absorption in the visible range and improved photocatalytic properties in hydrogen evolution experiments.[24] Elemental doping is achieved by the substitution of single atoms in the triazine or heptazine building unit by elements like O,[25] C,[26] P,[27] S,[28] B,[29,30] I,[31,32] F.[33,34] Also the introduction of metals like Fe into carbon nitrides has been reported by Wang et al. and describe another approach to effectively tune the photocatalytic properties of carbon nitrides.[35]

In this work, we focus on molecular doping of PTI and discuss a facile variation of a novel synthesis, that has been recently reported by the authors.[36] This approach is conducted in an aqueous solution of melamine, in which between two platinum electrodes a voltage is applied which exceeds the potential of the electrolysis of water. During the process of electrolysis radical species like hydroxyl or superoxide radicals are formed. These radicals react with the melamine precursor to form melamine radicals which polymerize.[37] In further experiments it was shown that hydroxyl radicals are expected to be the main species in this reaction.[36] Through step-wise substituting melamine by TAP in this approach it was possible to tune the optical properties of PTI leading to a significant improvement of the photocatalytic properties. For the degradation of methylene blue (MB) a significant increase of the degradation rate could be observed with an increasing amount of TAP.

**Results and Discussion**

In order to achieve molecular doping of PTI as it is shown in Figure 1 melamine was substituted by TAP. Therefore, solutions were prepared with 0, 25, 50, 75 and 100 mol% of TAP. The total molar amount of melamine and TAP was constant as well as the sodium hydroxide concentration. The reactions were conducted in a vessel shown in the photograph in Figure SI1. The reactions were followed by UV-Vis spectroscopy for the evaluation of the reaction times. In accordance to the Lambert-Beer law the reactions were considered to be completed when no further changes in the absorption spectra depicted a constant concentration of the product. These spectra are shown in Figure SI2 and it is shown that the reaction time increases with the amount of TAP used in synthesis. After precipitation using hydrochloric acid and a washing procedure, the product was freeze-dried and the incorporation was demonstrated by applying different characterization methods.

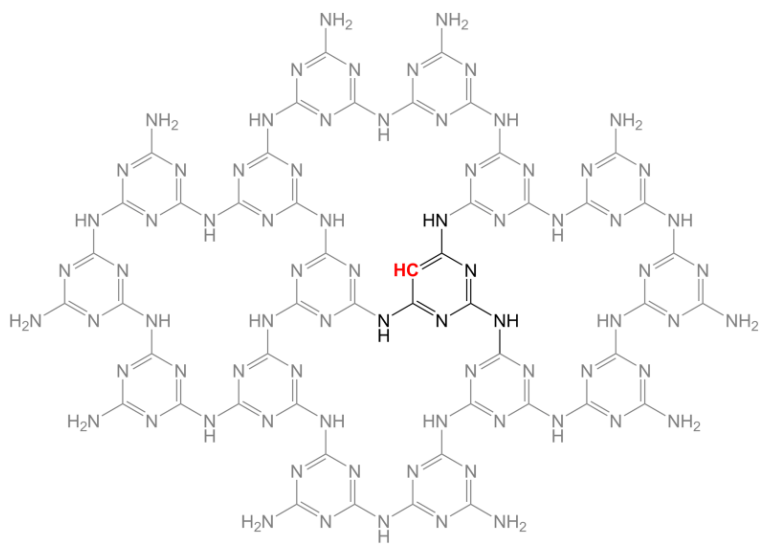

Figure 1     The structure of PTI is illustrated in grey. An *s*-triazine unit, which has been substituted by a pyrimidine unit, is shown in black. The exchange of nitrogen by carbon is marked red.

The powder X-ray diffraction (XRD) pattern is shown in Figure 2. It consists of two reflexes. The reflex at 27.75 ° is related to the (002) stacking of conjugated aromatic systems.[1] The calculated interlayer distance is found to be 3.22 Å. With an increasing amount of TAP substituting melamine in the synthesis the position is at a constant angle with an exception of the complete substituted sample, which shows a slight shift to smaller angles (27.59 °). This behavior has been observed for other carbon nitrides and is explained by the substitution of a nitrogen atom resulting in reducing the stability and structural regularity.[19,22] The increase of the full width half maximum (fwhm) indicates that the stacking of the single layer PTI is decreased and smaller crystal domains are formed. Using the Scherrer equation sizes of 19.1 nm are found for PTI synthesized using melamine whereas the complete substitution leads to sizes of 7.5 nm. This trend is also observed for the reflex at 10.6 °, which is labeled as (100) and is defined by an in-plane distance of 8.33 Å. It shows a slight shift towards larger angles with an increasing amount of TAP in the synthesis. If melamine is completely substituted, this reflex cannot be observed anymore, which has also been reported for carbon nitrides which are prepared by the thermal condensation of TAP or barbituric acid.[22] The fwhm of the (100) reflex and the calculated sizes decrease with increasing molecular doping from 10.5 nm (0 % TAP) to 6.7 nm (75 % TAP). All in all in XRD a decrease in the crystalline quality of the product can be observed in terms of the crystal size in the ab-plane as well as along the c-axis.

Figure 2 also shows the FTIR spectra of the products. The characteristic absorptions of carbon nitrides can be observed. The peak at 805 cm$^{-1}$ is defined by the out-of-plane bending mode of the triazine ring. The region between 1800 cm$^{-1}$ and 1100 cm$^{-1}$ shows various absorption maxima for imide and nitride stretching vibrations. At 1445 and 1260 cm$^{-1}$ the characteristic bands for the C=N and C-NH-C vibration are found, respectively.[9] The presence of oxygen containing functional groups like C=O or C-OH is indicated by the bands around 1730 cm$^{-1}$ and 1180 cm$^{-1}$ and is in correspondence with the results of the elemental analysis. The incorporation of oxygen has been reported in a previous work and has its origin in the synthesis approach.[36] The bands at 3350 cm$^{-1}$ and 3220 cm$^{-1}$ are related to NH-stretching vibrations. The peak at 2350 cm$^{-1}$ can be assigned to carbon dioxide. The FTIR spectra of the 0, 25, 50 and 75 % samples are similar apart from an increased broadening of the peaks that can be observed with an increasing TAP amount. This broadening indicates a lower structural order. The similarity of the spectra of TAP doped carbon nitrides was also recognized by Ho et al.[38] In contrast, the bands observed for PTI 1660, 1535, 1445 and 1380 cm$^{-1}$ disappear in the 100 % sample and two broad bands at 1633 and 1355 cm$^{-1}$ appear. The peak at 1260 cm$^{-1}$, which is assigned to the C-NH-C vibration, can be observed at low intensities for all of the samples.[9] The low intensity of this signal combined with the sizes calculated from the XRD pattern shows the low degree of polymerization of the product. The intensity of the out-of-plane bending mode of PTI, which is found at 805 cm$^{-1}$, decreases with increasing amount of TAP used in the synthesis and cannot be observed for the 100 % sample.

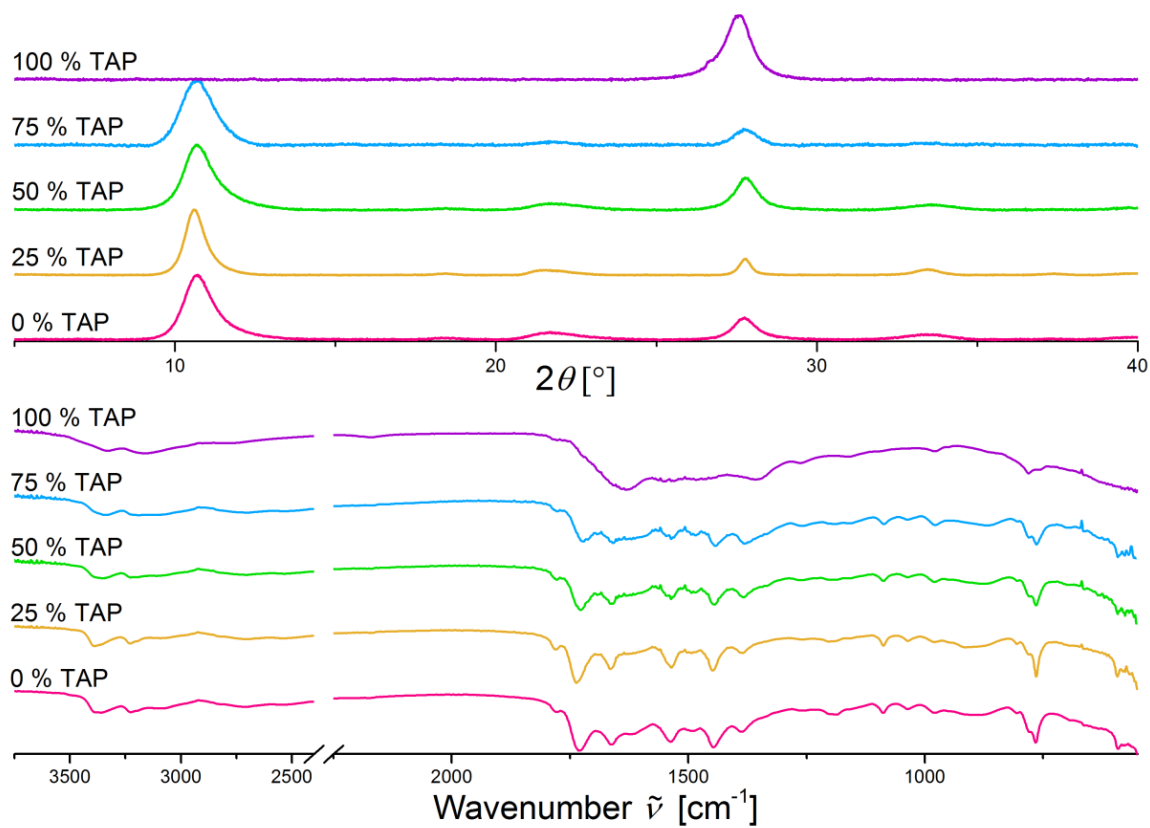

Figure 2  The XRD pattern of PTI shows two characteristic reflexes at 10.6 ° and 27.7 °, which are related to an in-plane periodicity (100) and interlayer stacking (002), respectively (top). FTIR spectrum spectra of TAP doped PTI. The vibration at 2350 cm$^{-1}$ which is contributed by $CO_2$ has been removed to improve the clarity (bottom).

The incorporation of TAP was further verified by elemental analysis. The amounts of carbon, nitrogen and hydrogen were determined by combustion analysis whereas the amount of oxygen was provided by pyrolysis analysis. The calculated sum formula of PTI is $C_3N_{4.5}H_{1.5}$. The values found for the samples with different molar percentages of TAP in the synthesis are presented in Table 1. The sum formula calculated for 100 % of TAP in the synthesis is $C_4N_{3.5}H_{2.5}$. Oxygen and the increased amount of hydrogen, which are present at a constant value in all of the samples, are explained by adsorbed water as well as by the incorporation of oxygen containing functional groups like C-OH and COOH due to the synthesis method.[36] The elemental analysis confirms the low degree of condensation of the as-formed product. The difference from the calculated and the measured nitrogen amount in the doped samples can be explained by terminating amine groups. In general, it can be shown that the C/N-ratio increases with the addition of TAP.

Table 1   Amounts of carbon, nitrogen, hydrogen and oxygen found in the sample by combustion and pyrolysis analysis.

|  | C mass% (atomic%) | N mass% (atomic%) | H mass% (atomic%) | O mass% (atomic%) | Sum Formula (integer carbon) |
|---|---|---|---|---|---|
| 0 % TAP | 26.39 (20.99) | 48.72 (33.23) | 3.61 (34.22) | 19.37 (11.57) | $C_3N_{4.75}O_{1.65}H_{4.89}$ |
| 25 % TAP | 27.74 (21.79) | 47.57 (32.04) | 3.74 (35.00) | 18.95 (11.17) | $C_3N_{4.41}O_{1.54}H_{4.82}$ |
| 50 % TAP | 28.43 (22.37) | 46.42 (31.31) | 3.73 (34.97) | 19.21 (11.35) | $C_3N_{4.20}O_{1.52}H_{4.69}$ |
| 75 % TAP | 28.52 (22.63) | 45.10 (30.69) | 3.70 (34.99) | 19.61 (11.68) | $C_3N_{4.07}O_{1.55}H_{4.63}$ |
| 100 % TAP | 30.94 (24.70) | 43.50 (29.78) | 3.53 (33.59) | 19.89 (11.92) | $C_3N_{3.62}O_{1.45}H_{4.08}$ |

In $^{13}C$ NMR spectra, which were recorded in deuterized DMSO, two peaks at 166 and 150 ppm are observed. These peaks can be assigned to the carbon next to the terminal $NH_2$ group and the carbon next to the bridging NH group. For the 0, 25 and 50 % TAP samples only these two peaks are observed. The 75 % TAP samples shows additional peaks in the region between 166 and 150 ppm that are related to the changes in the chemical environment by the carbon substituted nitrogen in the triazine ring. This additional carbon atom contributes to the spectrum by the peak at 73 ppm.[39] For the 100 % TAP samples only this peak at 73 ppm and the peak at 150 ppm can be observed. The $^{13}$C NMR spectra are shown in Figure SI3.

Figure 3A+B shows two photographs of the different samples in liquid and solid state as well as the spectra obtained by UV-Vis spectroscopy and photoluminescence measurements in solution (DMSO as solvent). The photographs show a colorless (solution) or pale yellowish color (solid) for the 0 % sample. With an increasing amount of TAP in the synthesis the samples turn yellow and orange. The 100 % sample appears brown. These observations are confirmed by the obtained spectra. The absorption spectra are defined by a shoulder at 410 nm, which becomes more pronounced with an increasing amount of TAP used in the synthesis. A shift of the absorbance in the visible range is clearly observable (Figure 3C). The solid state spectra and the corresponding Tauc plots show a main feature with a Tauc gap between 3.3 and 3.4 eV for the 0, 25, 50 and 75 % sample. For the 100 % sample this feature can be found at 3.15 eV. A second features appears at a lower energy of 2.7 eV and can be observed for the 25, 50, 75 and 100 % sample (Figure SI4). Thus, with an increasing amount of TAP an additional absorption process takes place at lower energies compared to the regular absorption of PTI. The normalized PL spectra show a broad peak at 425 nm for the product with 0 % of TAP in the synthesis (Figure 3D). This wavelength is slightly red-shifted compared to values reported in the literature.[40] The complete substitution of melamine by TAP results in a red shift of this peak to 564 nm. The PL spectra for the 25, 50 and 75 % samples consist of several peaks, whose positions can be reconstructed by multiplying the PL spectra of the 0 and 100 % samples. The red shift of the PL is in good agreement with the energy observed for the shoulder in the absorption spectra and the calculated Tauc gap.

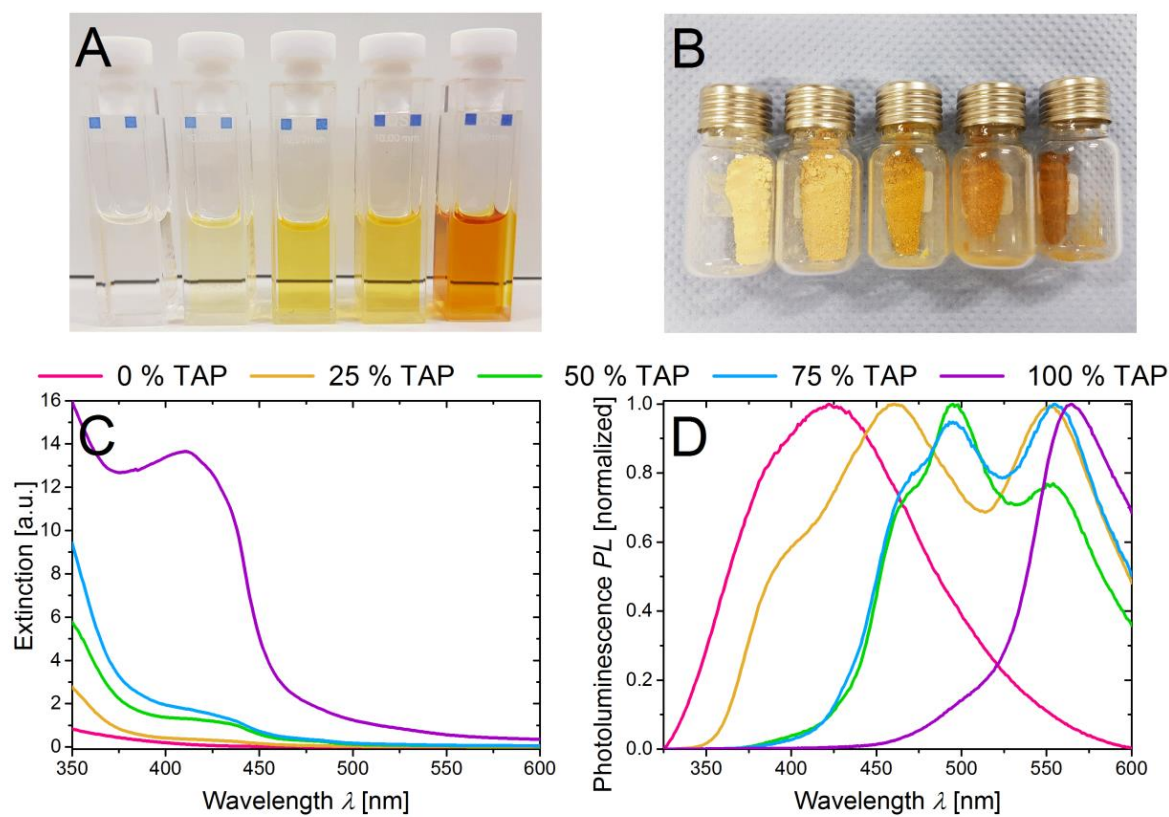

Figure 3    Photographs of the 0, 25, 50, 75 and 100 % samples (from left to right) in DMSO (A) and freeze-dried (B). UV-Vis spectra (C) and photoluminescence spectra (D) of the samples measured in DMSO.

SEM and TEM images, which are presented in Figure SI5, show no change in the morphology of the carbon nitride. The samples consist of loosely arranged networks made of less than 100 nm wide ribbon-like structures.

Prior to photocatalytic experiments the specific surface of the catalyst was determined by BET-measurements. For the 0, 25, 50, 75 and 100 % TAP samples specific surfaces of 51.4, 45.5, 42.5, 52.9 and 53.9 $m^2/g$ were found, respectively. The observed values are similar and do not show any trend which could be related to the amount of TAP used in the synthesis. Hence, changes in the photocatalytic properties of the different samples cannot be related to the specific surface.

**Photocatalytic experiments**

**Degradation of methylene blue**

All of the prepared samples were capable of decomposing methylene blue (MB) without the addition of any co-catalyst. The adsorption/desorption equilibrium was set after 140 min stirring in the dark (Figure SI6). The calculated adsorption capacities of the molar amount of MB per mass of the catalyst are similar for all of the samples (0.051±0.006 mmol/g) and are in good agreement with the PTI synthesized without sodium hydroxide (0.04 mmol/g).[36] The multiplication of the adsorption capacity by the Avogadro constant and the size of a MB molecule of 135 Å$^2$ leads to an estimated surface of 41.5 m$^2$/g.[41] This value is close to the specific surfaces found by BET measurements. Therefore, it is assumed that most of the surface is accessible by the MB molecules and a nearly complete coverage is achieved. The obtained spectra are shown in Figure SI7. The logarithmic plots of the intensities at the absorption maximum of MB are shown in Figure 4A. The evaluation of the observed degradation rates, which is depicted in Figure 4B, shows an increase depending on the amount of TAP used in the synthesis. The highest rate of $2.8 \times 10^{-3}$ min$^{-1}$ was observed for the sample, in which 75 % of melamine was substituted by TAP. We were able to boost the degradation rate by a factor of 4 in relation to the rate of $0.6 \times 10^{-3}$ min$^{-1}$ of the pure PTI sample. We explain the increased degradation rate by the enhanced overlap of the lamp spectrum and the absorption spectra of the different PTI samples (Figure SI8) as well as the different electronic properties resulting from the incorporation of TAP. In detail, Bhunia et al. performed DFT calculations and showed that TAP doping and the resulting incorporation of carbon results in a smaller bandgap by lowering the absolute potential of the valence band.[23] Nevertheless, the incorporation of melamine in a certain amount seems to be important as the sample prepared by 100% of TAP shows a lower degradation rate. For comparison melon was prepared by thermally induced polycondensation of melamine by standard procedures.[42] The evaluated rate of the degradation of MB using melamine derived melon is $0.4 \times 10^{-3}$ min$^{-1}$. It is shown, that PTI exhibits a slightly higher degradation rate compared to melon and by molecular doping an increase of the rate by a factor of 7 is achieved. The stability of the catalyst was evaluated for the 75 % TAP sample by the separation of the catalyst and the MB solution after every circle, subsequent washing with water and redispersing of the catalyst in a fresh MB solution. It was shown that the degradation rate decreases with every recycling step (Figure 4C). After 3 cycles the degradation rate decreases to 27.5 % of the initial value. This decrease cannot be explained by the loss of catalyst material during the washing procedure and may result from adsorbed degradation products that block the active sites of the catalyst. Additional washing steps might improve the recyclability. In addition, scavenger experiments were conducted to identify the active species in the degradation process (Figure 4D). *tert*-butanol (*t*-BuOH), *para*-benzoquinone (BQ) and EDTA-2Na have been added as scavengers for hydroxyl radicals, superoxide radicals and holes, respectively.[43,44] The addition of *t*-BuOH results in a slightly decreased degradation rate, whereas for BQ for the first hour no degradation of MB but a shift in the absorption of BQ is observed (Figure SI7I). This shift decreases during the 4 hours of degradation and the rate of the degradation of MB increases. We assume that evolving superoxide radicals react preferential with BQ. When most of the BQ molecules have reacted the MB degradation starts to take place. The observed rate is similar to the reference without any scavenger molecule. These results are in accordance to our previous work.[36] The addition of EDTA-2Na effects a small decrease of the degradation rate. Thus, it is concluded that superoxide radicals are the main active species of the degradation of MB followed by hydroxyl radicals and holes.

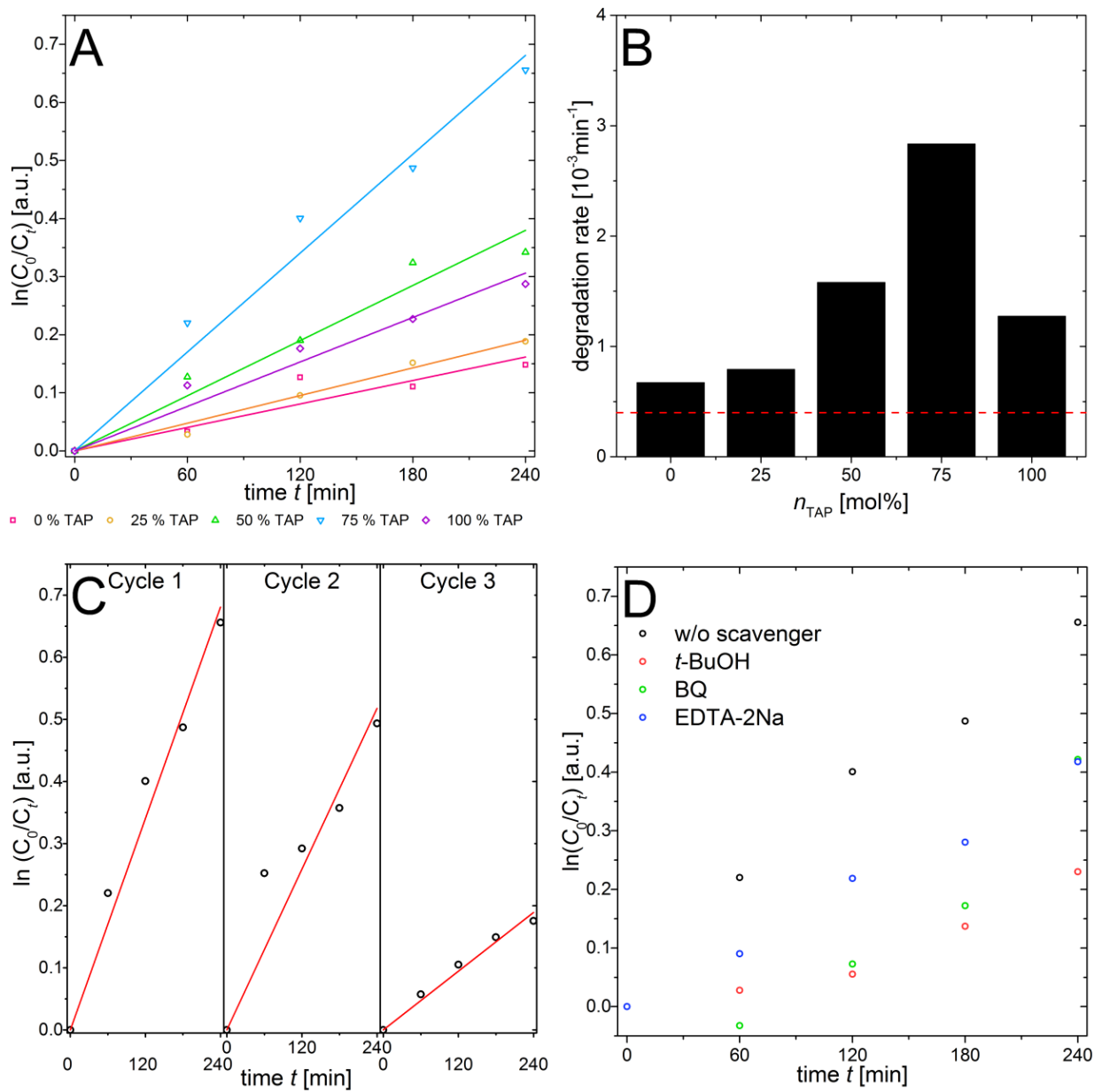

Figure 4    Logarithmic plot of the degradation of MB against time (A) and a comparison of the degradation rates of MB using catalysts with different TAP amounts in the synthesis and melon (dashed red line) (B). Recycling experiments of the 75 % TAP sample for three cycles (C) and scavenger experiments using *t*-BuOH, BQ and EDTA-2Na (D).

**Conclusion**

In summary, we report a facile modification of the electrochemical synthesis approach recently reported by the authors, which is based on the formation of radicals during electrolysis of water and the reaction of those with melamine.[36] Just by partially substituting melamine with TAP in the synthesis it was possible to further improve the photocatalytic properties of PTI. We were able to tune the molar ratio of melamine and TAP from 0 to 100 %. With an increasing amount of TAP the electronic structure of PTI has been tuned and a decrease of the bandgap could be observed. The smaller bandgap results in changes in the optical properties of the product in regard to a red-shift in the visible region that can be observed in absorption as well as in photoluminescence experiments. This red-shift enhanced the photocatalytic properties of PTI and in dye degradation experiments we were able to observe an increase of the degradation rate by a factor of 4 in relation to the undoped PTI or 7 if compared to melon prepared by thermally induced polycondensation of melamine.

**Experimental**

*Chemicals*

Melamine (99 %) was purchased from Sigma-Aldrich, 2,4,6-triaminopyrimidine (>97 %) was purchased from Sigma-Aldrich, sodium hydroxide (97 %) was purchased from Grüssing GmbH, hydrochloric acid (37 %) was purchased from VWR, dimethyl sulfoxide (99.7 %) was purchased from ACROS organics, methylene blue was purchased from Merck Chemicals GmbH, *tert*-butanol (99 %) was purchased from Grüssing GmbH, EDTA-2Na (>98.5 %) was purchased from Sigma-Aldrich, and *para*-benzoquinone was purchased from Sigma-Aldrich and recrystallized in 2-propanol.

*Synthesis of melon*

Melon was synthesized by thermal induced polycondensation of melamine. 5 g (39.6 mmol) of melamine were placed in a semi-closed crucible and heated to 550 °C with a ramp of 5 °C/min.

After 2 h at 550 °C the product was allowed to cool down naturally, milled into a powder using mortar and pestle, washed with 100 mL boiling water and dried in a vacuum furnace at 60 °C.

*Synthesis of poly (triazine imide)*

The synthesis is similar to the synthesis for poly (triazine imide) that has been recently reported.[36] 630 mg (5 mmol) of melamine and 800 mg (20 mmol) of sodium hydroxide were dissolved in 100 mL of demineralized water at room temperature under vigorous stirring. In a three electrode set up with a reference electrode (Ag/AgCl/KCl), counter electrode (Platinum) and working electrode (Platinum) a voltage of 5 V was applied for 90 min at room temperature under continuous stirring. The distance between counter and working electrode was 0.5 cm. The current was in the range of 600 mA. A photograph of the set-up is shown in Figure SI1.

The product was obtained in solution. By addition of hydrochloric acid the pH value was adjusted to obtain a suspension. The product was separated by centrifugation for 5 min at 7500 rpm (13206 rcf) and washed twice with 30 mL hot demineralized water.

2,4,6-Triaminopyrimidine was used for molecular doping. The synthesis was as described above, but melamine was substituted by 0 mol%, 25 mol%, 50 mol%, 75 mol% and 100 mol% of 2,4,6-triaminopyrimidine. Additionally, the duration of the reaction was changed to 120 min, 210 min, 240 min, 330 min and 300 min, respectively.

*Photocatalytic experiments*

50 mg of the different catalyst materials were dispersed in 100 mL of a 10 mg/L methylene blue in water solution and stirred for 140 min in the dark to set an adsorption/desorption equilibrium. As a light source a 300 W Xenon lamp (UXL-302-O) combined with a water filter to avoid heating effects by infrared radiation was used. Aliquots of 4 mL were taken every 60 min. The aliquots were centrifuged once at 11000 rpm (28408 rcf) in order to remove the catalyst material. The supernatant was filtered using a hydrophilic syringe filter (nylon, 0.2 μm pore size).

For recycling experiments the catalyst was collected after every cycle and separated from the MB solution by centrifugation at 11000 rpm (28408 rcf) for 15 min at a temperature of 5 °C. The supernatant was discarded. The catalyst was redispersed in 30 mL of water and after the adsorption/desorption equilibrium was set again separated from the solution by centrifugation. This procedure was repeated twice. The second and third circle of the degradation was conducted as described above.

The reactive species in the degradation process were identified by adding *tert*-butanol (10 vol.%), EDTA-2Na (1 mmol) or *para*-benzoquinone (1 mmol) to the experiment.

*Methods*

UV-Vis-Spectroscopy: A quartz cuvette with an optical path length of 10 mm was used for absorption and photoluminescence experiments. UV-Vis absorption spectra were obtained with a PerkinElmer Lambda 25 spectrometer in a two-beam set-up. Photoluminescence spectra were obtained with a Horiba Fluoromax-4 spectrometer.

Fourier-transform infrared spectroscopy: The FTIR spectra were collected on a Varian 660 infrared spectrometer with an attenuated total reflection sampling technique. The set-up was purged with nitrogen for 1 h. 32 scans were used to collect the spectrum.

X-Ray diffraction: Powder X-Ray diffraction measurements were carried out on a Philips X'Perts PRO MPD diffractometer with monochromatic X-Ray radiation from a copper anode with a wavelength of 1.54 Å (CuKα). The samples were dispersed in water, drop casted and dried on a <911> or <711> silicon wafer.

Electronmicroscopy: SEM images were collected with a Zeiss MA10 electron microscope with a LaB$_6$ emitter at 5 kV. TEM images were collected on a JEOL-1011 with a thermal emitter operated at an acceleration voltage of 100 kV.

Gas adsorption: The Brunauer-Emmett-Teller (BET) surface area was obtained by nitrogen sorption experiments using a Quantachrome Autosorb iQ Automated Gas Sorption System at 77 K.

NMR spectroscopy: $^{13}$C NMR spectra were recorded in DMSO in a $^{13}$C{$^{1}$H}-BB experiment using 3072 scans on a Bruker AVANCE 400 MHz spectrometer.

**Associated Content**

**Supporting Information**

The Supporting Information is available free of charge on the ACS Publications website at DOI: .

Photograph of the experimental set-up, additional UV-Vis spectra and Tauc plots, $^{13}$C NMR spectra, SEM and TEM images, UV-Vis spectra of aliquots taken during dye degradation, and xenon lamp spectrum.

**Acknowledgement**

L.H. and C.K. gratefully acknowledge financial support of the European Research Council via the ERC Starting Grant "2D-SYNETRA" (Seventh Framework Program FP7, Project: 304980). C.K. thanks the German Research Foundation DFG for financial support in the frame of the Cluster of Excellence "Center of ultrafast imaging CUI" and the Heisenberg scholarship KL 1453/9-2.

**Author Information**

**Corresponding author**

*E-mail: christian.klinke@swansea.ac.uk

**ORCID**

Leonard Heymann: 0000-0001-6177-6943

Christian Klinke: 0000-0001-8558-7389

**Notes**

The authors declare no competing financial interest.